\documentclass{llncs}
\usepackage{cite}
\usepackage{graphicx}
\usepackage{amsfonts}
\usepackage[cmex10]{amsmath}
\usepackage{euscript}
\usepackage{pstricks}
\usepackage{color}
\usepackage{flushend}
\usepackage{enumerate}
\usepackage[shortlabels]{enumitem}
\usepackage{subfig}
\usepackage{todonotes}
\usepackage{changes}

\newtheorem{heuristic}{Heuristic}

\newcommand{\preset}[1]{\ensuremath{{}^\bullet {#1}}}
\newcommand{\postset}[1]{\ensuremath{{#1}{}^\bullet}}
\newcommand{\Bag}[1]{\ensuremath{Bag({#1})}}

\begin{document}
\title{Verification of Reachability Problems for Time Basic Petri Nets}
\author{Matteo Camilli}
\institute{Dept. of Computer Science \\ Universit\`a degli Studi di Milano, Italy \\ \email{camilli@di.unimi.it}}
\maketitle

\normalem

\begin{abstract}
Time-Basic Petri nets, is a powerful formalism for modeling real-time systems where time constraints are expressed through time functions of marking's time description associated with transition, representing possible firing times.
We introduce a technique for reachability analysis based on the building of finite contraction
of the infinite state space associated with such a models.
The technique constructs a finite symbolic reachability graph relying on a sort of time coverage,
and overcomes the limitations of the existing available analyzers for Time-Basic nets,
based in turn on a time-bounded inspection of a (possibly infinite) reachability-tree.
A key feature of the technique is the introduction of the Time Anonymous concept, which allows the identification
of components not influencing the evolution of a model.
A running example is used throughout the paper to sketch the symbolic graph construction.
The graph construction algorithm has been automated by a \textsc{Java} tool-set,
described in the paper together with its main functionality and analysis capability.
A use case describing a real-world example has been employed to benchmark the technique and the tool-set.
The main outcome of this test are also presented in the paper.
%Ongoing work, in the perspective of integrating with a model-checking engine,
%is shortly discussed.
\end{abstract}

\section{Introduction}
\label{sec:intro}
Time-Basic (TB) Petri nets \cite{UnifiedWay91} belong to the category of nets in which system time constraints
are expressed as numerical intervals associated to each transition, representing
possible firing instants, computed since transition's enabling time.
Tokens atomically produced by the firing of a transition are thereby associated to time-stamps
with values ranging over a determined set.
With respect to the well-known representative of this category, i.e.,
Time Petri nets \cite{Berthomieu91}, interval bounds in TB nets are linear functions
of timestamps in the enabling marking, rather than simply numerical constants.
TB nets thus represent a much more expressive formal model for real-time systems. 
The reachability analysis of TB nets is still recognized as an open problem \cite{Hudak2010}.
Available analysis techniques and tools (e.g., \cite{Hudak2010,Cabernet93}) are based on inspecting
a finite portion of the potentially infinite reachability-tree generated by a TB net.
But for particular cases, only time-bounded properties can be inferred from
TB net's state-space exploration by using this kind of analyzers. 
The technique described in this paper tries to overcome this major limitation.
It relies on a symbolic reachability graph algorithm,
which is in turn based on a relative notion of time
and on a symbolic state definition in which
variables are used instead of numerical time-stamp values, and
time dependencies are expressed by linear constraints.
The core of the algorithm is a procedure verifying inclusion between symbolic states, that relies in turn on 
two key concepts: the \emph{erasure of absolute times} and the \emph{identification of anonymous timestamps}.
Broadly speaking, the erasure of absolute times allow us to identify equality/inclusion relationships among states
although they have a diverse displacement with respect to the initial time.
The anonymous timestamp concept relies on the fact that there may exist components for which
timestamp values can be ignored, as not influencing the evolution of the model.
The procedure permits in many cases to build a sort of \emph{time coverage} finite reachability graph.
This paper represents an extended version of \cite{Bellettini11}, which take a deeper look 
at the anonymous timestamp concept and introduces all the adopted heuristics able to find this kind of components.

The symbolic graph construction, including the search of time anonymous timestamps, has been automated by a tool-set written
in \textsc{Java}.
The output is a structure enriched with information on edges which might be exploited
during property evaluation. The tool-set currently includes a module for
the automatic verification of reachability properties expressed as conditions on markings.
%Differently from Time Petri net's analysers, that implement a (symbolic) reachability graph algorithm \cite{} based on a concept of state-class, the algorithm implemented in the tool-set
%%, in particular the technique adopted for eliminating unessential time dependencies,
%produces as output a kind of \textit{(time-)coverage graph}.
%%% questa parte potrebbe essere espansa per fare un confronto un poco pi\`u ampio
%{\color{red}@@@ DA RIVEDERE
%Due to its modular structure, the tool-set may be used as stand-alone module,
%or it may be integrated in two graphical Petri net editors:
%%the original environment for TB nets,
%the \texttt{Cabernet} package, and the \texttt{PIPE}(2) open-source package~\cite{pipe2}. That makes it possible to
%exploit consolidated structural analysis algorithms for the verification of the untimed part of TB nets.
%}
As use case we'll use the gas burner example, that is widely used in literature as a representative of a small real system. A complete and formal description can be found in \cite{Ravn93}, and the corresponding  TB net model was introduced in \cite{IPTES-PDM41}. An excerpt will be used as running example to explain in a rather informal
way the essential points of symbolic graph construction. Only some relevant new core definitions  are formally given.

\section{Time Basic Nets}
\label{sec:tbnets}
%%%%
Time Basic nets are Petri nets where each token is associated with a time-stamp representing the instant at which it has been created. The domain of timestamps is 
$\mathbb{R}^+$. The structure of a Time Basic net is a triplet $(P,T,F)$, where $P$ and $T$ are finite
sets, called \emph{places} and \emph{transitions}, respectively, s.t. $P \cap T = \emptyset$,
and $F$ is the \emph{flow relation}, $F \subseteq (P \times T) \cup (T \times P)$.
Let $v \in P \cup T$: $\preset{v}$, $\postset{v}$ denote the backward and forward adjacent sets
of $v$ according to $F$, respectively, also called pre/post-sets of $v$.
A (time-stamp) \emph{tuple} of $t \in T$ is an association $en:\, \preset{t} \rightarrow \mathbb{R}^+$. 
Each transition $t$ is associated with a \emph{time function} $f_t$ which maps a tuple $en$
of $t$ to a (possibly empty) set of $\mathbb{R}^+$ values. A \emph{marking} (state)
is a mapping $m:\,P \rightarrow \Bag{\mathbb{R}^+}$, $\Bag{A}$ being the set of multiset over $A$.
A tuple $en$ of $t$ is said to be \emph{enabling} in $m$, in accordance to a \emph{weak} semantics
(as explained next), if $\forall p \in \preset{t}$ $en(p) \in m(p)$
and $f_t(en) \neq \emptyset$. $f_t(en)$ represents the possible firing times for $en$.
Letting $en$ be an enabling tuple of $t$ in $m$, a pair $(en,\tau)$, $\tau \in f_t(en)$, is said 
a \emph{firing instance} of $t$ (in $m$).
The firing of $(en,\tau)$ produces the new marking $m'$, s.t.
$\forall p \in {^{\bullet}t} \setminus {t^{\bullet}}$ $m'(p) = m(p) - en(p)$,
$\forall p \in {t^{\bullet}} \setminus {^{\bullet}t}$ $m'(p) = m(p) + \tau$,
$\forall p \in {t^{\bullet}} \cap {^{\bullet}t}$ $m'(p) = m(p) - en(p) + \tau$;
for all remaining places, $m'(p) = m(p)$.
This will be as usual denoted $m [(en, \tau)> m'$.

Hereafter a time function $f_t$ is defined by a pair of linear functions
$[lb_t, ub_t]$, denoting parametric interval bounds.
$lb_t, ub_t$ are in turn formally expressed in terms of
(a non empty set of) places in $\preset{t}$: $lb_t(en)$, $ub_t(en)$
are the numerical expressions obtained by replacing each place occurrence
$p$ with $en(p)$.
Time-functions must be monotonic, i.e.,
%the set of time-stamps associated with a tuple $en$ cannot contain a timestamp less than
$\forall en$ $lb_t(en) \geq$
$enab$ $\equiv$ $max(\{en(p)\}, p\in \preset{t})$.
We will keep such assumption implicit in their formal notations.
%SEMANTICA

The set of firing times $f_t(en)$ can be interpreted in at least two different ways, leading to different time semantics for \emph{each} transition $t$. A first interpretation states that an enabling tuple $en$ of $t$ \emph{can} fire at any instant $\tau \in f_t(en)$. Transitions with one
such semantics are referred to as \emph{weak}. A second interpretation states that an enabling tuple \emph{must} fire at an instant $\tau \in f_t(en)$, unless it is disabled by the firing of any conflicting enabling tuple at an instant no greater than the latest firing time of $t$. 
Transitions with one such semantics are referred to as \emph{strong}. Thereby the enabling condition previously given must take into account also the possible presence of other strong enabling tuples \cite{UnifiedWay91}.
Notice that the only possible semantics for Time Petri Nets \cite{Berthomieu91} is strong.

%MONOTONICITA' SCATTI a LIVELLO RETE
In order to meet an intuitive notion of time, TB net firing sequences are restricted to the set of firing sequences whose firing times are monotonically non decreasing with respect to the firing occurrences.
%NON ZENONICITA'
However, the time of a firing may be equal to the enabling time of the tuple that belongs to the firing. Intuitively this means that an effect (the firing) can occur with no delay after the cause (that  enables it) is fulfilled. Therefore, it is possible to have sequences of firings where the time does not change. In practice, it is useful to restrict the attention to a subclass of TB nets, such that there exist no infinitely long firing sequences which take a finite amount of time (non Zenonicity).
%%%%

\begin{figure*}[ht]
\begin{center}
{\includegraphics[width=1\textwidth]{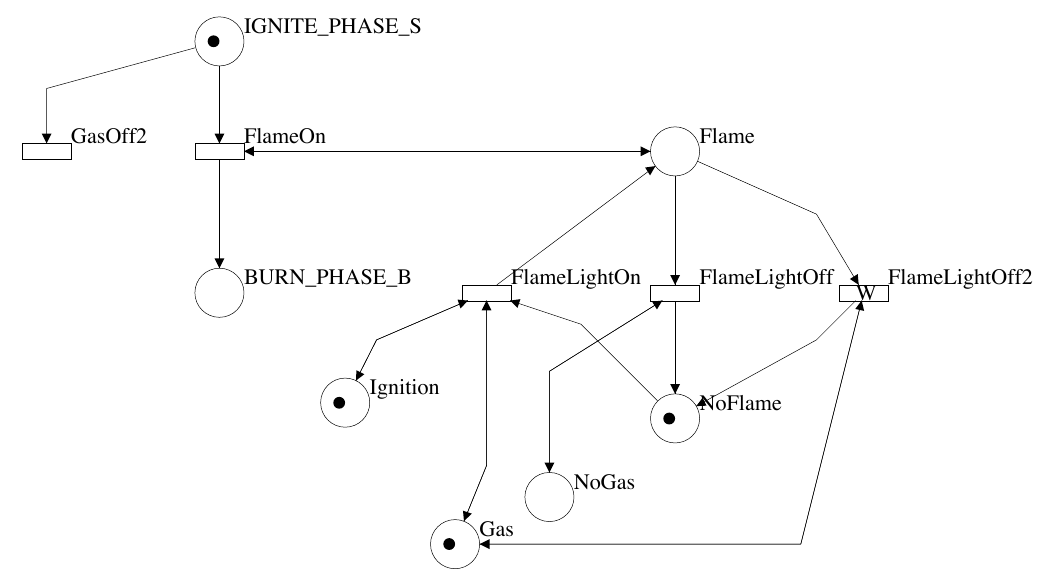}}
{\small
\vskip 0.5cm
{\scriptsize
\begin{tabular}{@{} llll @{}}
Initial marking &  \multicolumn{3}{l}{$IGNITE\_PHASE\_S\{T_0\}\,\,Ignition\{T_0\}\,\,Gas\{T_0\}\,\,NoFlame\{T_0\}$}\\
Initial constraint & $0 \le T_0 \le 10$ \\
 & \\
%\bf{FlameLightOff} & ($max(Flame,NoGas) , NoGas + 0.1$) \\
%\bf{FlameLightOff2} & ($max(Flame,Gas), max(Flame,Gas)+100$) with weak time semantic \\
\bf{FlameOn} & \multicolumn{3}{l}{[$IGNITE\_PHASE\_S + 0.01 , max(\{Flame+0.1 , IGNITE\_PHASE\_S + 0.01\})$] }\\
\bf{FlameLightOn} & [$enab+0.5 , enab+0.5$]  & \bf{FlameLightOff} & [$enab , NoGas + 0.1$] \\
\bf{GasOff2} & [$enab + 2 , enab + 2$] & \bf{FlameLightOff2} & [$enab, enab+100$] weak time semantic \\
\end{tabular}
}
}
\caption{Running example.}
\label{fig:example}
\end{center}
\end{figure*}

Consider the excerpt from the use case, depicted in Fig.~\ref{fig:example}.
It relates to the \emph{Ignite Phase}, just after the ignition transformer has been started and the gas valve has been opened. In this phase the controller must check if the flame has been lighted within a specific deadline, otherwise a recovery procedure that brings the system to \emph{Idle} has to be activated.
All transitions are strong, but \emph{FlameLightOff2}. This permits us to express the $possibility$ that an event occurs within a given time interval.

The flame turns on if there are \emph{Ignition} and \emph{Gas} (transition \emph{FlameLigthOn}), but it can turn off if no gas is supplied (transition \emph{FlameLigthOff}) or due to a failure, caused e.g. by wind (transition \emph{FlameLigthOff2}).
The time function associated with transition \emph{FlameOn} (representing the system passing to $burn state$ after recognizing that the flame has turned on)
%is denoted by
%$$[IGNITE\_PHASE\_S + 0.01 , max(Flame+0.1 , IGNITE\_PHASE\_S + 0.01)]$$
can be interpreted as follows: $FlameOn$ cannot fire before 0.01 time units elapse since the appearance of a token
in place $IGNITE\_PHASE\_S$ (the minimum permanence time in $ignite state$) and implicitly not before the timestamp in place $Flame$. 
The firing time cannot exceed the maximum between the timestamp of the token in place $IGNITE\_PHASE\_S$ plus 0.01 time units and the time-stamp of the token in place \emph{Flame} plus 0.1 (i.e., the system recognizes the presence of a flame within this 0.1 units).
Noticeably, this is an example of constraint that cannot be directly expressed using Time Petri Nets formalism \cite{Berthomieu91}. 

\section{Time coverage reachability analysis}
\label{sec:technique}
The analysis technique presented in this paper extends the capability of the existing analyzer for TB nets \cite{Merlot93},
which uniquely permits the verification of bounded invariance and response properties,
through the inspection of a time-bounded symbolic reachability tree generated from a TB net.
%The new tool implements some ideas sketched in \cite{articoloConfrontoAnalisi}.

The new technique aims at building a finite graph instead of an infinite tree
for a wide category of TB nets. A combination of three complementary
ideas is exploited. First, symbolic states are compared to check subset relationships.
For that purpose, using a consolidated approach, timestamp symbols no more
occurring in the marking description are eliminated
from the linear constraint associated to a symbolic state,
%of a TB net
independently of how it has been reached.
% Questa estensione porta alla costruzioni di grafi aciclici, anche se ci sono 2 eccezioni:
% cicli a tempo zero (esclusi pero' in precedenza) e marcatura senza limite superiore... 
% T0 >=2   che puo' includere successive
Identifying subset relations between generated symbolic states (markings plus constraints),
is necessary for recognizing cyclic paths, but it is not enough in many situations.
As time progresses, periodic occurrences of equivalent conditions may be unrecognizable
simply due to their different offsets with respect to system's time zero.
This observation leads us dealing with the second aspect.
%Second,  i
In the very common case a TB model contains no reference to \textit{absolute times} (i.e., not as offset respect to enabling timestamps)
in transition time functions, it is possible to remove any references
to the ``absolute zero'' from symbolic states.
This permits a periodic equivalent behavior to be recognized.
%conseguenze su espressività degli stati. Certe proprietà non possono essere immediatamente verificate.
The cost is a lossy information about state displacement along absolute time.
We'll discuss this aspects in section~\ref{sec:prop}. 
Let us only point out that this kind of information could be recovered, if necessary,
in a second step by retracing only the path(s)
leading to the state of interest, or (at least partially) by combining the information on edges.
The third key feature of the technique is the introduction of the \emph{time anonymous} (\emph{TA}) concept. This relates to the fact that in a symbolic state there may exist tokens
whose timestamp values can be \emph{forgotten}, as not influencing the evolution of a model. 
Several heuristics have been implemented, based on a mix of structural and state-dependent patterns,
each characterizing one such situation. 
This enhances the ability of merging states, and permits facing situations where the presence of
dead tokens could reintroduce a sort of \emph{symbolic} absolute zero, nullifying the achievements
at the previous points. 
Again, the cost to pay is a minor loss of information, as discussed later.
There is some resemblance with the approach used in the construction of (topological) coverage graphs:
the missing information is the exact timestamp of tokens instead of their exact number.
\emph{TA} recognition might be also exploited to introduce a topological notion of coverage
for TB nets (section~\ref{sec:conclusion}).  

\subsection{Basic notions}
\label{sec:graphconstr}
In order to understand the rationale behind the symbolic reachability graph construction technique for TB nets,
we shall use once again the running example in Fig.~\ref{fig:example}.
Let us only introduce a few basic notions used in the sequel, referring to \cite{Ghezzi94}
(where the symbolic reachability tree for TB nets is defined) for a full formalization.

Let $TS = \{T_i\}$, $i \geq 0$, be the set of time-stamp symbols.
A \emph{symbolic state} $S$ is a pair $\langle M, C\rangle$, where $M:\,P \rightarrow \Bag{TS}$,
$C$ is a (satisfiable) constraint formed by linear inequalities involving $TS$ symbols
occurring in $M$ (so called symbolic marking).

Unless otherwise specified, we shall refer to a \emph{normal form}:
if  $k$ different $TS$ symbols occur in $M$,
they are $T_0,\ldots,T_{k-1}$, such that  $\forall i: 0\ldots k-2$, $C \Rightarrow T_i \leq T_{i+1}$.

An ordinary marking $m$ is represented by $S:\,\langle M, C\rangle$ if and only if $m$ is obtained from $M$ by a numerical replacement $\sigma:\,TS \rightarrow \mathbb{R}^+$, $\sigma$ being a solution of $C$. We say that $S$ is contained in $S'$ ($S \subseteq S'$) if and only if the corresponding represented ordinary markings are.  

A mapping $en_s\,:\, \preset{t} \rightarrow TS$ is said a \emph{symbolic
tuple} of $t$. The notation $(en_s,t)$ will be sometimes used.
%Let $en_s^{-1}(\tau)$ = $\{ p\}$, $en(p)=\tau$.
%$en_s$ will be formally denoted by a tuple of symbols.
The \textit{symbolic evaluation} of a time function $f_t$, denoted $f_t(en_s)$,
is obtained by replacing each occurrence of $p \in \preset{t}$ in the formal expressions $lb_t$,
$ub_t$, with $\tau = en_s(p)$.
%(such an association will also be denoted $p:\,\tau$, depending on the context).
  
According to a (monotonic) weak time semantics, 
$(en_s, t)$ is said a \emph{symbolic enabling} 
in $S$ if $\forall p \in \preset{t}$ $en_s(p) \in M(p)$ and $C'$: $C \wedge lb_t(en_s) \leq T_{k} \leq ub_t(en_s) \wedge
T_{k-1} \leq T_{k} $ is satisfiable, i.e., there exists at least one numerical
substitution (tuple) $en$ for $en_s$ that makes $C$ satisfiable and $f_t(en)$ non empty.
As already said the symbolic enabling condition 
is a bit more complex to take into account strong enablings: an example will be provided in Sect.~\ref{sec:time-cov-graph}.  

The firing of a symbolic enabling $(en_s, t)$ produces the new symbolic state $S' : \langle M', C'\rangle$, where $M'$ is obtained from $M$ by removing $en_s(p)$ from each place $p \in {^{\bullet}t}$, and putting the new symbol $T_{k}$ in all places in
${t^{\bullet}}$, in full analogy with the ordinary firing rule.
That is denoted $M [(en_s, t)> M'$.
$S'$ represents all the possible ordinary markings reachable from any marking
represented by $S,$ by means of any firing instance corresponding to $(en_s, t)$.

\subsection{Time-coverage graph construction}
\label{sec:time-cov-graph}
\begin{figure*}[ht]
\centering
\includegraphics[width=1\textwidth]{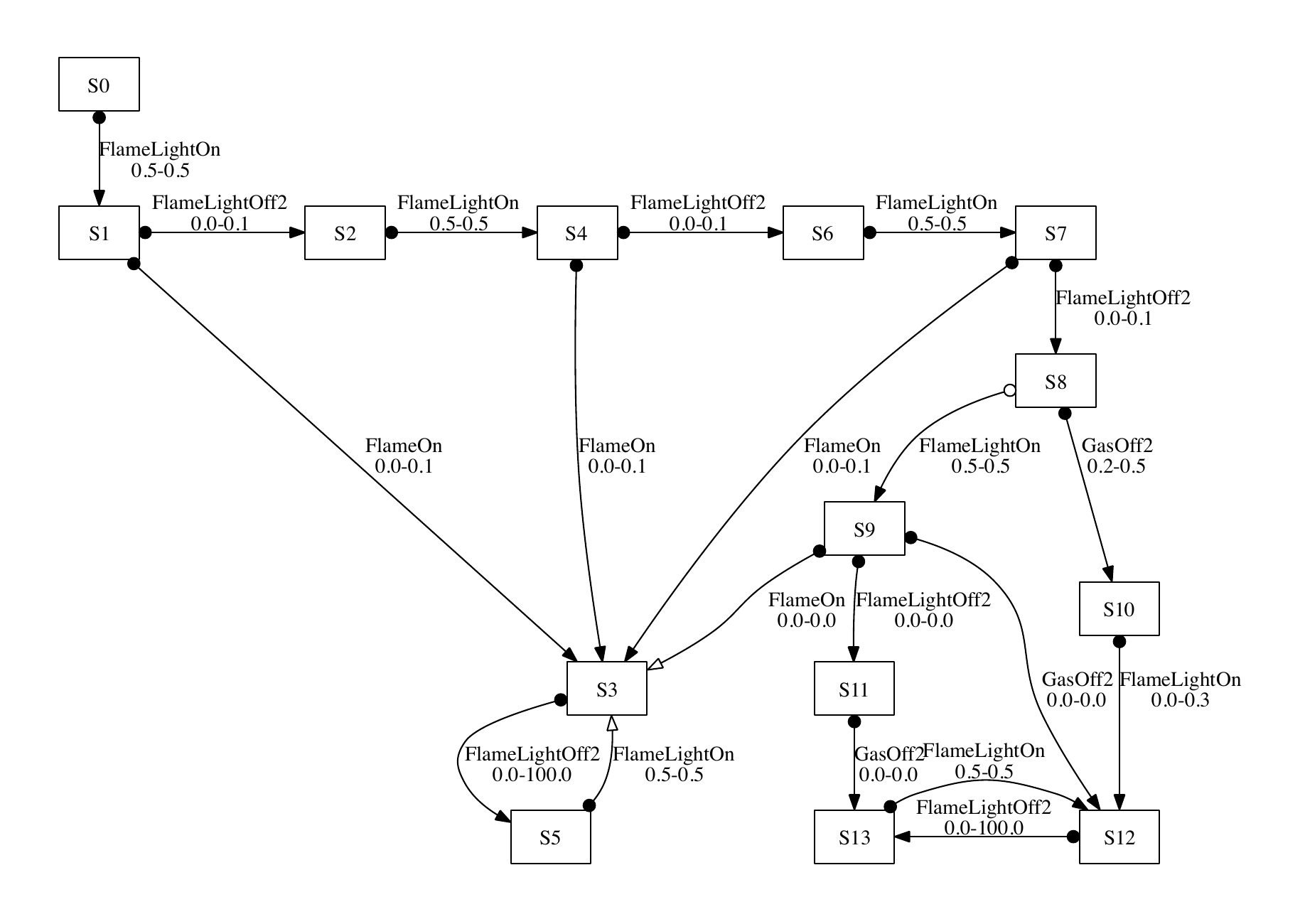}
\caption{Sample reachability graph.}
\label{fig:graph}
\end{figure*}

The time-coverage symbolic reachability graph generated by the running example, composed by 14 symbolic states,
is presented in Fig.~\ref{fig:graph}.\footnote{This picture has been automatically obtained by using GraphViz visualization software~\cite{graphviz} on the output generated from the tool-set.}

%STATO S8
%"mark" = "Gas{T1} IGNITE_PHASE_S{T0} Ignition{TA} NoFlame{TA}";
%"const" = "TL-T0-1.5>=0 && TL-T0-1.8<=0 && TL-T1==0";
%ci sono due transizioni abilitate
%GasOff2   e FlameLightOn
%GasOff2 scatta esattamente 2 unità di tempo dopo IGNITE....  
%T0+2
%FlmaeLightOn... scatta esattamente a 0.5 istanti di tempo dopo il max tra Gas e Ignition
%Gas+0.5
%T1+0.5
%T0+2<T1+0.5 (a causa semantica forte) solo per sottoinsieme
%t0+1.5 < T1
%T1=t0+1.5 puo' scattare FlameLON
%altrimenti deve scattare GasOff2
%T0+1.5 <= T1 <= T0+1.8
%
%
%
%senza TA e con tempi assoluti
%STATO S12 
%"mark"= "Gas{T2;}IGNITE_PHASE_S{T0;}Ignition{T1;}NoFlame{T2;}"
%"const"= "T0-10.0<=0 && T1-T0-1.5>=0 && T1-T0-1.7<=0 && T2-T1>=0 && T2-T1-0.1<=0"];
%con TA con tempi assoluti
%node: S12
%"mark" = "Gas{T1;}IGNITE_PHASE_S{T0;}Ignition{TA;}NoFlame{TA;}";
%"const" = "T0-10.0<=0 && T1-T0-1.5>=0 && T1-T0-1.8<=0";
%senza TA e con tempi relativi
%node: S10
%"mark" = "Gas{T2;}IGNITE_PHASE_S{T0;}Ignition{T1;}NoFlame{T2;}";
%"const" = "T1-T0-1.5>=0 && T1-T0-1.7<=0 && T2-T1>=0 && T2-T1-0.1<=0";
The adopted notation for states is:
a square for symbolic states, a double square for symbolic states containing some deadlocks.
Concerning edges (i.e., symbolic enablings), the format of head and tail specifies the kind of relation between source and
target.

The normal case
is black head and tail, e.g., from $S0$ to $S1$:
considering any marking represented by $S0$ it is always possible to follow that edge and to reach all the markings represented by $S1$.
 
Let us consider the symbolic state $S8$, formally described as follows:
$$ \left.\begin{array}{lll}M8 &: & Gas\{T_1\} \,\, IGNITE\_PHASE\_S \{T_0\}
\\ & & Ignition\{\textit{TA}\} \,\, NoFlame\{\textit{TA}\}
\\C8 &: & T_1 \ge T_0+1.5 \wedge T_1 \le T_0+1.8
\end{array}\right.$$
We can observe that, with respect to the original definition of symbolic state, a first extra time-stamp symbol
is present, \emph{TA} (time anonymous). This new symbol can occur only on the marking.
Postponing an intuitive explanation of when and how symbol \emph{TA} is
introduced in a symbolic state representation,
we can think of it as a token carrying on an unspecified time-stamp,
which has been shown unessential for the computation of transition firing times.
    
The ``candidates'' for symbolic enabling in $S8$ are: 
\begin{itemize}
\item $(\langle T_0\rangle, \textit{GasOff2})$ 
\item $(\langle \textit{TA},T_1,\textit{TA}\rangle, FlameLightOn)$.
\end{itemize}
Firing times are computed by (symbolically) evaluating transition time functions, as explained above.
For \emph{GasOff2} the (only) inferred firing time is $\{T_0+2\}$.
Time function evaluation is slightly different for \emph{FlameLightOn}, due to the occurrence
of \emph{TA} in the pre-set tuple: this symbol is \textit{erased} (Definition~\ref{erasure} in the following section) during symbolic
evaluation: $enab = max(\{ \textit{TA},T_1,\textit{TA}\}) \equiv max(\{T_1\}) = T_1$.
The inferred firing time in this case is $\{T_1+0.5\}$.

Since both transitions have a strong semantics, there are two additional constraints specifying that
the firing time of one cannot be greater than the (maximum) firing time of the other.
%\footnote{The set
%of firing times of a (strong) transition in $S$ also depends on the enablings of the other strong transitions.}
They are $C_{\textrm{\tiny{GO2}}}:\, T_0+2 <= T_1+0.5$ and $C_{\textrm{\tiny{FLO}}}:\, T_1+0.5 <= T_0+2$, respectively.

Since both $C8 \wedge C_{\textrm{\tiny{GO2}}} \wedge T_2$ = $T_0+2$ and $C8 \wedge C_{\textrm{\tiny{FLO}}} \wedge T_2$ = $T_1+0.5$ are satisfiable, $(\langle T_0\rangle, \textit{GasOff2})$ and $(\langle \textit{TA},T_1,\textit{TA}\rangle, FlameLightOn)$
are in fact symbolic enablings in $S8$.
It is important to note that $C8 \Rightarrow C_{\textrm{\tiny{GO2}}} \wedge T_2 = T_0+2$, i.e.,  all the markings represented by $S8$ enable the transition \emph{GasOff2}. Instead $C8 \not\Rightarrow C_{\textrm{\tiny{FLO}}} \wedge T_2 = T_1+0.5$, i.e., only a subset of the markings expressed by $S8$ enable the transition  \emph{FlameLightOn}. This is highlighted in the graph by the white tail of the edge from $S8$ to $S9$.

Consider now the firing of $(\langle T_0\rangle, \textit{GasOff2})$: it only consumes tokens.
In such cases the symbolic firing rule slightly differs from the original one.
A second special symbol, $TL$ (Time Last), is introduced.
$TL$ can occur only on the constraint of a symbolic state and has an intuitive meaning:
it stands for the last firing time of the TB net and it permits
a correct interpretation of the model's time semantics.\footnote{In this paper, when $TL$ is left implicit,
it coincides with the ``last'' generated timestamp $T_k$.} 
The reached symbolic state $S10$ is:
$$ \left.\begin{array}{lll}M10 &: & Gas\{T_1\} \,\,  Ignition\{\textit{TA}\} \,\, NoFlame\{\textit{TA}\} \\C10 &: & C8 \wedge T_2 = T_0+2 \wedge TL=T_2
\end{array}\right.$$
The normalization step eliminates symbols $T_2$ (the symbolic firing time) and $T_0$, as they occur only in $C10$, instead it leaves symbol $TL$. That results in (after a timestamp renaming):  
$$ \left.\begin{array}{lll}M10 &: & Gas\{T_0\} \,\,  Ignition\{\textit{TA}\} \,\, NoFlame\{\textit{TA}\} \\C10 &: & TL \geq  T_0+0.2 \wedge TL \leq T_0+0.5
\end{array}\right.$$
Another circumstance that causes the introduction of $TL$ symbol in a symbolic state representation is when
the maximum timestamp symbol $T_k$ is replaced with \emph{TA}. The identification of a Time Anonymous in a given symbolic state is the next topic we treat. 

The graph in Fig.~\ref{fig:graph} contains two looping paths: between states $S3$ and $S5$, and between $S12$ and $S13$ respectively. That happens because in the extrapolated sub-model (Fig.~\ref{fig:example}), no expected actions are activated after the system exits the \emph{ignition phase} (e.g., closing the gas valve in the event of fail, or stopping ignition), so that an unbounded sequence of \emph{FlameLightOff2};\emph{FlameLightOn} is possible. 

The white head of the edge from $S5$ to $S3$ means that at least one of the ordinary
markings represented by $S3$ is not reachable by following that edge. 
This happens when a newly built symbolic state is recognized to be strictly contained in an existing one.
What permits recognizing inclusion between states in this specific case
is the usage of \emph{Time Anonymous} timestamps (Definition~\ref{def:TA-repl}). $S3$ is formally defined as:
$$ \left.\begin{array}{lll}M3 &: & Gas\{\textit{TA}\} \,\, BURN\_PHASE\_B \{\textit{TA}\}
\\ & & Ignition\{T_0\} \,\, Flame\{T_1\}
\\C3 &: &  T_1 \ge T_0 \wedge T_1 \le T_0+0.1
\end{array}\right.$$
Without using \emph{TA}s, its original definition ($S3'$) would be:
$$ \left.\begin{array}{lll}M3' &: & Gas\{T_0\} \,\, BURN\_PHASE\_B \{T_1\}
\\ & & Ignition\{T_0\} \,\, Flame\{T_1\}
\\C3' &: & T_1 \ge T_0 \wedge T_1 \le T_0+0.1
\end{array}\right.$$
Let us figure out what would be the model evolution from $S3'$, without introducing \emph{TA}.
After the firing sequence \emph{FlameLightOff2};\emph{FlameLightOn}\footnote{We omit in this description symbolic enablings, the TB net being safe.} a state $S3''$ would be reached, defined in turn as:
$$ \left.\begin{array}{lll}M3'' &: & Gas\{T_1\} \,\, BURN\_PHASE\_B \{T_0\}
\\ & & Ignition\{T_1\} \,\, Flame\{T_1\}
\\C3'' &: & T_1 \ge T_0 +0.5 \wedge T_1 \le T_0+100.5
\end{array}\right.$$
Since $S3'' \not\subseteq S3'$ and $S3' \not\subseteq S3''$, there is no possibility to merge them and in fact the analysis tool would produce an infinite firing sequence.

Back to $S3$, we note it corresponds to $S3'$ but for holding \emph{TA} symbols in places $BURN\_PHASE\_B$ and $Gas$ instead of $T_1$ and
$T_0$, respectively.
Token $T_1$ in $BURN\_PHASE\_B$ however is not (and will never be) involved in any symbolic enabling because $BURN\_PHASE\_B$ has an empty postset (Heuristic 0 in the following section), so it is immediately marked as \emph{TA}.
Token $T_0$ in $Gas$ instead is in the preset of transitions \emph{FlameLightOn} and \emph{FlameLightOff2}. As for \emph{FlameLightOn}, the tokens in place $Ignition$ and in place $Gas$ carry on the same timestamp, so either of them is enough to correctly evaluate transition's time function. As for \emph{FlameLightOff2}, the token in place $Gas$ carries on redundant information due to the simultaneous presence of $T_1$ in \emph{Flame}, that superseded it (Heuristic~\ref{heu-2}).

$S3''$ seems really different from $S3$, but nearly the same heuristics permits us to replace $T_0:\,BURN\_PHASE\_B$ ($T_i:\,p$ denotes the occurrence of a timestamp in a place) and $T_1:\,Gas$ with \emph{TA}s. That eliminates all the occurrences of $T_0$
from the marking. After timestamp renaming, we obtain the normal form:
$$ \left.\begin{array}{lll}M3'' &: & Gas\{\textit{TA}\} \,\, BURN\_PHASE\_B \{\textit{TA}\}
\\ & & Ignition\{T_0\} \,\, Flame\{T_0\}
\\C3'' &: & true
\end{array}\right.$$
However there is still a difference  with respect to $S3$:
places \emph{Ignition} and \emph{Flame} hold the same timestamp, but this boils
down to a condition already represented by $S3$ ($T_1 = T_0 \Rightarrow C3$), so $S3''$ is recognized as a state contained in $S3$.

%{\color{red}@@@
Notice that the other cycle on the graph, between $S12$ and $S13$,
is due to the adoption of a relative notion of time, i.e.,
it does not depend on the introduced \emph{TA} concept.    
%}

%TIME LIMIT
An important setting of the legacy tool \cite{Cabernet93} was the \emph{time limit}, a positive interval time that guaranteed the finiteness of the symbolic reachability tree of a TB net. 
Upon elimination of absolute time references it has been substituted by a \emph{relative time limit}.
%As said, the analysis technique cannot guarantee a resulting finite reachability graph.
This positive interval specifies the maximum admissible distance between different timestamps in a state,
and allows one to deal with possibly infinite reachability graph.
The tool-set checks whether a symbolic state includes any ordinary states for which the distance between $TL$ and $T_0$ (the oldest meaningful timestamp) exceeds the time limit, marking that state as \emph{not to be expanded}.
The rationale behind is that reaching such a user defined limit might be a symptom of the presence of
unrecognized ``dead tokens'', reintroducing absolute time references.
If we analyzed the running example disabling \emph{TA} recognition,
the resulting graph would be infinite, unless a time limit is set.
For example, setting this limit to 3 (time units), 25 symbolic states would be generated:
13 already included in the presented graph,
the others corresponding to a partial unrolling of the loop between $S3$ and $S5$.
%@@@ questo era per dire che volendo uno puo' scegliere di non usare TA e provare a mettere un time limit in modo da non perdere le  informazioni. Una altra opzione a livello di tool potrebbe essere quella di vietare la trasformazione in TA di gettoni in particolari posti.}

%ARCHI
The output generated by the tool-set associates 
a couple of numerical values to edges of the graph,
corresponding to the minimum and maximum time distances from the source node to the target node.
This permits us to partially recover time relations between nodes that were lost due to
the removal of absolute times references from constraints.
In the following section we'll show how to exploit them.

\section{Time Anonymous}
The notion of time anonymous relies on the fact that in a symbolic state there may exist tokens
whose timestamp values can be \emph{forgotten}, as not influencing the evolution of a model.
The adopted symbol to denote a time anonymous timestamp is \emph{TA}, and it represents an undefined 
time value in the past chosen between the initial time and the time limit \emph{TL}.
The \emph{TA replacement} task (formally defined in the next section) allow us to build, in many cases, a finite 
reachability graph. In fact, the presence of ``dead'' tokens in a model, i.e. those tokens that cannot be consumed
by firing transitions, reintroduce a sort of \emph{initial time} that would prevent the discovery of equality/inclusion
relationships among states.

\begin{figure}[htbf]
\centering
\includegraphics[width=.5\columnwidth]{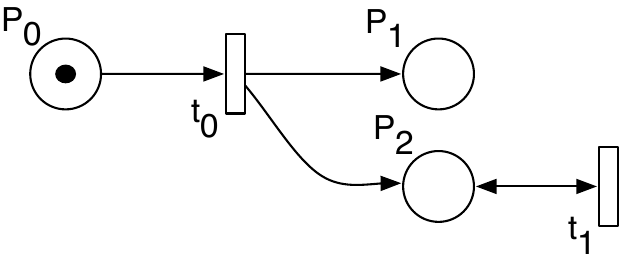}
\caption{Simple TB net example generating a ``dead'' token.}
{\small
\vskip 0.5cm
{\scriptsize
\begin{tabular}{@{} llll @{}}
\bf{Initial marking} &  $P_0\{T_0\}$ \\
\bf{Initial constraint} & $0 \le T_0 \le 1$ \\
 \end{tabular}
 \\
 \begin{tabular}{@{} llll @{}}
\bf{$t_0$} & [$enab + 0.2 , enab + 0.3$] \\
\bf{$t_1$} & [$enab + 0.5 , enab + 0.7$] \\
\end{tabular}
}}
\label{fig:pn-ta}
\end{figure}

\begin{figure*}[htp]
  \centering
  \caption{Infinite (a) and finite (b) representations of the reachability graphs extracted from the model shown in Fig.~\ref{fig:pn-ta}.}
  \subfloat[Reachability graph without $TA$ replacement.]{\label{fig:pn-ta-graph-1}\includegraphics[width=.9\columnwidth]{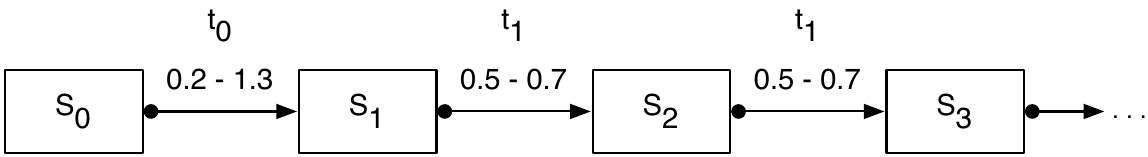}}
  \\
  %\vspace{0.8cm}
  \subfloat[Reachability graph with $TA$ replacement.]{\label{fig:pn-ta-graph-2}\includegraphics[width=.55\columnwidth]{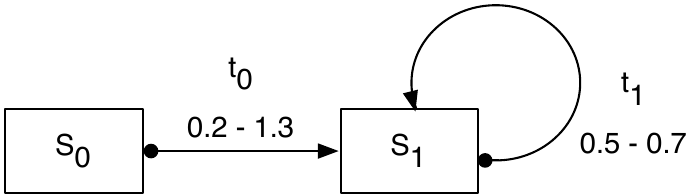}}
\label{fig:graphs}
\end{figure*}

As a simple example, let us consider the model described in Fig.~\ref{fig:pn-ta}.
Transition $t_0$ is enabled in the time lapse $[T_0 + 0.2, T_0 + 0.3]$. Its firing produces two new tokens,
respectively into $P_1$ and $P_2$ with a timestamp $T_1$ representing a value chosen in such a time interval.
This new configuration enables $t_1$ which can fire infinitely many times, by consuming and immediately after
creating a token in $P_2$, each time with a new timestamp.
Although the erasure of absolute times, the presence of a ``dead'' token in $P_2$, creates a sort of time marker
which would make the reachability graph infinite, as we can see in Fig.~\ref{fig:pn-ta-graph-1}.

After the initial state $S_0$, reachable states are all equal in terms of symbolic marking: $P_1\{T_0\} P_2\{T_1\}$ but they have 
different  constraints: 
\begin{itemize}
\item $C_{S_1} = 0.2 \le T_0 \le 1.3 \wedge T_0 + 0.5 \le T_1 \le T_0 + 0.7$
\item $C_{S_2} = 0.2 \le T_0 \le 1.3 \wedge T_0 + 1.0 \le T_1 \le T_0 + 1.4$
\item $C_{S_3} = 0.2 \le T_0 \le 1.3 \wedge T_0 + 1.5 \le T_1 \le T_0 + 2.1$
\end{itemize}
and so forth, departing $T_1$ from $T_0$ further and further.
Anyway, it is worth noting that $T_0$ does not influence the evolution of the model, thus we can forget about 
this value replacing it with an anonymous timestamp $TA$. The $TA$ replacement cause the erasure of $T_0$
from constraints enabling the identification of equality relationships among states.
In fact, a $TA$ timestamp does not have any relationships with other symbolic values because it represents any
time value in the past.
Therefore, all the states after the initial one, would have the same constraint: $C_{S_1} = TRUE$.
The finite reachability graph, resulting from the analysis of Fig.~\ref{fig:pn-ta}, using $TA$ replacements, is shown in Fig.~\ref{fig:pn-ta-graph-2}.

We identified three different typologies of tokens disclosing a negligible symbolic time:
\begin{itemize}
\item The first category is composed of ``dead'' tokens. A token $t_k$ is dead if belongs  to a place with an empty postset. 
Therefore such a token will be never consumed by firing transitions. It is possible to statically identify places that may
contain dead tokens. 
\item The second category contains all tokens $t_k$ such that $t_k$ belongs to a place $p$ with a non empty postset, 
and $t_k$ cannot be consumed by firing transitions. I.e. foreach $t \in p^{\bullet}$, any symbolic tuple $(en_s,t)$, such that $en_s(p) = t_k$ is not an symbolic enabling.
It is not possible to statically evaluate places containing such a tokens.
\item This latter category regards all tokens $t_k$ such that $t_k$ can be consumed by a firing transition,
but its firing time is not evaluated in terms of the timestamp associated with $t_k$. As the previous category, 
we must search for such a tokens dynamically, during the graph construction.
\end{itemize}

It is worth noting that, a symbolic enabling $(en_s, t)$ such that $lb_t(en_s)=TA$ makes the lower bound
 $lb_t(en_s)$ equals
to $TL$, in fact a $TA$ lower bound means that $TL$ exceeds the minimum enabling time.
Anyway, in case the preset of a transition $t$ contains only ``$TA$ tokens'', $t$ cannot fire because
both the lower bound and the upper bound of $t_f$ would be any time value in the past, thus
we cannot determine whether it represents an empty set.
The reason of a $TA$ replacement of all tokens belonging to ${^{\bullet}t}$ could be that
foreach symbolic tuple $(en_s,t)$, $TL > ub_t(en_s)$. Thus, if such a tokens does not contribute to
the evaluation of 
possible firing times of other transitions,  we can forget about all their symbolic times.

The next section introduces a formal definition of a ``TA replacement'' and all the adopted heuristics 
in order to find time anonymous timestamps during the graph building.

\section{Formal Definitions}
Let us formalize some core concepts previously outlined, focusing in particular on \emph{TA} and coverage.
For the sake of 
%space and 
readability, definitions involving transitions refer to the weak semantics.
\begin{definition}[symbolic state]
\label{def:symb-state-weak}
A symbolic state $S$ is a pair $\langle M, C\rangle$, where $M$ is a function $P \rightarrow \mathbf{Bag}(TS \cup \{\text{TA}\})$, and $C$ is a (satisfiable) linear constraint defined on $TS_{\text{M}} \cup \{\text{TL}\}$, $TS_{\text{M}} \subset TS$ being the finite set of symbols $T_i$ occurring on $M$,
such that $\forall T_i \in TS_{\text{M}}$, $C \Rightarrow TL \geq T_i$.
\end{definition}

%%%%%
\begin{definition}[well-defined erasure]
\label{erasure}
Let $g_t$ be the formal expression of a linear function. The \textit{erasure}
of a set of symbols $E \subset \preset{t}$ from $g_t$, denoted ${g_t}_{[\neg E]}$, is well-defined 
if it doesn't violate the arity of any operators occurring in $g_t$.
\end{definition} 
\noindent Consider for instance $t$, s.t. $\preset{t} = \{p_1,p_2\}$, and $f_t \,:$ $[max(\{p1,p2\}),p2+0.5]$,
where, $max\,:$ $2^{\mathbb{R}^+}\setminus \emptyset \rightarrow \mathbb{R}^+$, 
$+\,:\,\mathbb{R}^+,\mathbb{R}^+\rightarrow \mathbb{R}^+$. Then, the erasure ${f_t}_{[\neg \{p_1 \}]}$ is well-defined and
results in $[p2, p2+0.5]$, instead ${f_t}_{[\neg \{p_2 \}]}$ is not well-defined.

A symbolic instance of $t$ is a mapping $en_s\,:\, \preset{t} \rightarrow TS \cup \{\textit{TA}\}$.
\\Let $en_s^{-1}(\tau)$ = $\{ p\}$, $en(p)=\tau$.

\begin{definition}[symbolic enabling]
\label{def:symb-enab-weak}
$(en_s, t)$ is said a symbolic enabling  in \\$S = \langle M, C\rangle$ if and only if:
\begin{enumerate}[i]
 \item $\forall p \in {^{\bullet}t}$, $en_s(p) \in M(p)$ 
 \item ${f_t}_{[\neg en_s^{-1}(\text{TA})]}$ is well-defined
 \item
%$C'\, :\, C \wedge lb_t(en_s)_{[\neg \text{TA}]} \leq T_{k} \leq ub_t(en_s)_{[\neg \text{TA}]} \wedge T_{k-1} \leq T_{k}$
$C \wedge {lb_t}_{[\neg en_s^{-1}(\text{TA})]}(en_s)  \leq {ub_t}_{[\neg en_s^{-1}(\text{TA})]}(en_s)$
 is satisfiable
\end{enumerate}
\end{definition}

Let $C \setminus X$ denotes the constraint obtained by eliminating variable $X$ from $C$, in such a way that
the solutions of $C \setminus X$ are ``projections'' of the solutions of $C$.
\begin{definition}[symbolic firing]
\label{def:symb-fir-weak}
Let $(en_s, t)$ be a symbolic enabling in $S = \langle M, C\rangle$, $k = |TS_{\text{M}}|$.
The firing of $(en_s, t)$ produces the new symbolic state $S' : \langle M', C'\rangle$, where
\begin{itemize}
\item $\forall p \in {^{\bullet}t} \setminus {t^{\bullet}}$, $M'(p) = M(p) - en_s(p)$
\item $\forall p \in {t^{\bullet}} \setminus {^{\bullet}t}$, $M'(p) = M(p) + T_k$
\item $\forall p \in {t^{\bullet}} \cap {^{\bullet}t}$, $M'(p) = M(p) - en_s(p) + T_k$
\item for all remaining places, $M'(p) = M(p)$
\item $C' = \, C \setminus TL \wedge {lb_t}_{[\neg en_s^{-1}(\text{TA})]}(en_s) \leq T_k \wedge T_k \leq {ub_t}_{[\neg en_s^{-1}(\text{TA})]}(en_s)
\wedge T_k \geq T_{k-1} \wedge TL = T_k$
\end{itemize}
\end{definition}
\noindent $C'$ may contain some symbols $T_i$ that have been withdrawn from $M'$. After eliminating
redundant variables, and (possibly) renaming left symbols, the reached state meets definition
\ref{def:symb-state-weak} and is in normal form.
%%%%%%

Let $\mathbf{R}(S)$ be the set of symbolic states reachable from $S$
\begin{definition}[valid \emph{TA}-replacement]
\label{def:TA-repl}
Given a state $S$, a timestamp occurrence $T_i : p$
is replaceable with \emph{TA}~$: p$ if and only if for each $S' = \langle M', C'\rangle\ \in \mathbf{R}(S)$
in which token $T_i : p$ is left (modulo timestamp renaming),
%such that $T_i:\,p$ is left (up to a timestamp renaming),
for each symbolic enabling $(en_s, t)$ in $S'$
%such that $p \in ^{\bullet}t$
s.t. $ en_s(p) = T_i$, ${f_t}_{[\neg \{p\}]}$ is a well-defined erasure and
\begin{center}
$C' \wedge max(\{TL,lb_t(en_s)\}) \leq ub_t(en_s) \Leftrightarrow C' \wedge max(\{TL,{lb_t}_{[\neg \{p\}]}(en_s)\}) \leq {ub_t}_{[\neg \{p\}]}(en_s)$
%$fs_t(en_s,EN(S'))$ (the constraint formally expressing the set of firing times for $en_s$)
%${fs_t}_{[\neg T_i:\,p]}(en_s,EN(S'))$, where ${fs_t}_{[\neg T_i:\,p]}$ is
%obtained by a symbolic evaluation of transition time functions,
%in which $p$ is \textit{erased} instead of being replaced by $T_i$.
\end{center}
\end{definition}

The new semantics of a symbolic state is provided by the following coverage notion. 

\begin{definition}[symbolic state coverage]
\label{state-cov}
Let $S$ = $\langle M, C\rangle$ be a symbolic state.
An ordinary marking $m$ is covered by $S$ if and only if
it corresponds to a numerical substitution $\sigma$ of symbols occurring in $M$, s.t. $\sigma$ satisfies $C$ and
%for each $S' = \langle M', C'\rangle\ \in \mathbf{R}(S)$,
for each ordinary enabling $en$ of $t$ in $m$,
%in $\mathbf{R}(m_{\sigma})$,
for each symbolic tuple $(en_s,t)$ in $S$ s.t. $en$ is a numerical substitution of $en_s$,
\begin{itemize}
\item ${lb_t}_{[\neg en_s^{-1}(\text{TA})]}$, ${ub_t}_{[\neg en_s^{-1}(\text{TA})]}$ are well defined
\item ${lb_t}_{[\neg en_s^{-1}(\text{TA})]}(en) = {lb_t}(en) \wedge {ub_t}_{[\neg en_s^{-1}(\text{TA})]}(en) = {ub_t}(en)$
\end{itemize}
\end{definition}

%\noindent A symbolic state is said empty if and only if its coverage is.
The next lemma sets the relationship between ordinary and symbolic instances (state transitions).

\begin{lemma}
\label{lem:cover}
Let $m$ be covered by $S$. If $m [(en, \tau)> m'$, then there exists a symbolic enabling $en_s$,
s.t. $en$ is a numerical substitution of $en_s$, $S [(en_s , t) > S'$ and $m'$ is covered by $S'$
\end{lemma} 

Let us finally report all the heuristics implemented by the tool to identify the \emph{TA} replacements 
commented in the previous sections. 
 
 Formally, a valid replacement of a timestamp occurrence
$T_i : p$ with \emph{TA}~$: p$, in $S$ = $\langle M, C\rangle$, according to definition \ref{def:TA-repl},
takes place whenever at least one of the following heuristic, is verified foreach $t \in p^{\bullet}$.
Note that if $p^{\bullet} = \emptyset$ (Heuristic 0), this condition is trivially true. 

%\begin{Heu}\label{heu-1}
%$\postset{p} = \emptyset$
%\end{Heu}

%\begin{Heu}\label{heu-2} $\forall t \in \postset{p}, ~ \bigwedge_{i=1}^{10}H(i)$
%\setlist[enumerate,1]{leftmargin=1.5cm}
\begin{heuristic}\label{heu-1}
  $\forall p' \in {^{\bullet}t}, \ M(p') \neq \emptyset$ \\
  \indent $\wedge \ f_t$ is in the form $[enab+c, enab+c']$ \\
  \indent $\wedge \ \exists p'\in \preset{t}$ $(\forall T_j \in M(p')\; C \Rightarrow T_j \geq T_i)$

\vspace{.2cm}
\noindent \emph{All places belonging to $^{\bullet}t$ are marked, $f_t$ is in the form $[enab+c, enab+c']$, but there exist 
another place containing only newer tokens. Thus tokens belonging to $p$ won't be used to compute the
enabling time.}
\end{heuristic}

\begin{heuristic}\label{heu-2}
 $\forall p' \in {^{\bullet}t}, \ M(p') \neq \emptyset$ \\
  \indent $\wedge \ f_t$ does not contain $p$ \\
  \indent $\wedge \ f_t$ does not contain $enab$ 
  
\vspace{.2cm}
\noindent \emph{All places belonging to $^{\bullet}t$ are marked,
but $p$ will not be used to compute possible firing times of $f$ because $f_t$ does not contain either the variable 
$p$or $enab$.}
\end{heuristic}

\begin{heuristic}\label{heu-3}
 $\forall p' \in {^{\bullet}t}, \ M(p') \neq \emptyset$ \\
  \indent $\wedge \ f_t$ is in the form $[max(\ldots)+c, max(\ldots)+c']$ \\
  \indent $\wedge \ \forall (en_s,t)$ symbolic enabling, ${lb_t}_{[\neg \{p\}]}(en_s) = lb_t(en_s) \wedge {ub_t}_{[\neg \{p\}]}(en_s) = ub_t(en_s)$ 
  
\vspace{.2cm}
\noindent \emph{All places belonging to $^{\bullet}t$ are marked, $f_t$ is in the form $[max(\ldots)+c, max(\ldots)+c']$, 
but foreach enabling tuple $en_s$, $f_t(en_s)$ equals $f_t[\neg \{p\}](en_s)$ (well defined erasure).
Thus neither $lb_t(en_s)$
nor $ub_t(en_s)$ refers to $T_i$.}
\end{heuristic}
  
\begin{heuristic}\label{heu-4}
 $\forall p' \in {^{\bullet}t}, \ M(p') \neq \emptyset$ \\
  \indent $\wedge \ \forall (en_s,t)$ symbolic enabling, $C \Rightarrow (TL > ub_t(en_s) \wedge TL \ge lb_t(en_s))$
  
\vspace{.2cm}
\noindent \emph{All places belonging to $^{\bullet}t$ are marked,
but $t$ is not enabled ($TL > ub_t(en_s)$) and tokens in $p$ won't be used to compute the lower bound of $f_t$ even if $t$ would be re-enabled by other tokens ($TL \ge lb_t(en_s))$).}
\end{heuristic}

\begin{heuristic}\label{heu-5}
 $\forall p' \in {^{\bullet}t}, \ M(p') \neq \emptyset$ \\
  \indent $\wedge \ \forall (en_s,t)$ symbolic enabling, $C \Rightarrow (lb_t(en_s) > ub_t(en_s) \wedge (TL \ge {lb_t}(en_s) \vee {lb_t}_{[\neg p]}(en_s) = lb_t(en_s) ))$
  
\vspace{.2cm}
\noindent \emph{All places belonging to $^{\bullet}t$ are marked,
but $t$ is not enabled ($lb_t(en_s) > ub_t(en_s)$) and tokens in $p$ won't be used to compute the lower bound of $f_t$ even if $t$ would be re-enabled by other tokens, in fact $TL \ge lb_t(en_s))$ or $p$ does not contribute to the 
evaluation of $lb_t(en_s)$.}
\end{heuristic}

\begin{heuristic}\label{heu-6}
 $\exists p' \in {^{\bullet}t}: \ M(p') = \emptyset$ \\
  \indent $\wedge \ f_t$ does not contain $p$
  
\vspace{.2cm}
\noindent \emph{$t$ is disabled in $S$ and $p$ does not contribute to the evaluation of $f_t$ foreach possible future
symbolic enabling.}
\end{heuristic}
  
\begin{heuristic}\label{heu-7}
 $\exists p' \in {^{\bullet}t}: \ M(p') = \emptyset$ \\
  \indent $\wedge \ {lb_t}$ contains $p$ \\
  \indent $\wedge \ {ub_t}$ does not contain $p$ \\
  \indent $\wedge \ \forall en \ s.t. \ (en, t)$ future symbolic enabling, $C \Rightarrow TL \ge lb_t(en)$

\vspace{.2cm} 
\noindent \emph{$t$ is disabled in $S$, $ub_t$ does not contain the variable $p$, and foreach possible future
symbolic enabling $(en_s,t)$, the lower bound $lb_s(en_s)$ will be greater or equal to $TL$.}
\end{heuristic}
  
\begin{heuristic}\label{heu-8}
 $\exists p' \in {^{\bullet}t}: \ M(p') = \emptyset$ \\
  \indent $\wedge \ f_t$ is in the form $[max(\ldots)+c, max(\ldots)+c']$ \\
  \indent $\wedge \ \forall en \ s.t. \ (en, t)$ future symbolic enabling, \\
  \indent \indent ${lb_t}_{[\neg \{p\}]}(en) = lb_t(en) \wedge {ub_t}_{[\neg \{p\}]}(en) = ub_t(en)$
\end{heuristic}
  
\begin{heuristic}\label{heu-9}
 $\exists p' \in {^{\bullet}t}: \ M(p') = \emptyset$ \\
  \indent $\wedge \ \forall en \ s.t. \ (en, t)$ future symbolic enabling, \\
  \indent \indent $C \Rightarrow (TL > ub_t(en) \wedge TL \ge lb_t(en))$
  \end{heuristic}
  
\begin{heuristic}\label{heu-10}
 $\exists p' \in {^{\bullet}t}: \ M(p') = \emptyset$ \\
  \indent $\wedge \ \forall en \ s.t. \ (en, t)$ future symbolic enabling, \\
  \indent \indent $C \Rightarrow (lb_t(en) > ub_t(en) \wedge TL \ge lb_t(en))$
\end{heuristic}
  
\vspace{.2cm}
Heuristics \ref{heu-8}, \ref{heu-9}, \ref{heu-10} are respectively conceptually similar to \ref{heu-3}, \ref{heu-4},
\ref{heu-5} except they refer to future symbolic enablings, being $t$ disabled within $S$.

\begin{heuristic}\label{heu-11}
Given a place $p'$ and a symbolic tuple $en_s$, let $\phi_{TA}(en_s,p')$ be a new symbolic tuple such that:
\[ \phi_{TA}(en_s,p')(p) = \left\{ 
  \begin{array}{l l}
    en_s(p) & \quad \text{if $p = p'$}\\
    {TA} & \quad \text{otherwise}
  \end{array} \right.\]
  
$\forall (en_s,t)$ symbolic enabling,
\\ \indent $C \Rightarrow (TL > ub_t(\phi_{TA}(en_s,p)) \wedge TL \ge lb_t(\phi_{TA}(en_s,p)))$

\vspace{.2cm}
\noindent \emph{This heuristic assesses whether the symbolic time $T_i$ influences the evaluation
%treats every single token one by one trying to understand their contribution in the evaluation 
$f_t(en_s)$. To this end, we consider $T_i$ as the last produced token by replacing each timestamp of $en_s$, except $T_i$, with a $TA$. If $T_i$ does not contribute to evaluate $f_t(en_s)$, even if this condition holds, we can replace it with a $TA$ timestamp.}
\end{heuristic}

%\noindent ($\tau \in M(p)$ means that $\tau$ occurs on multiset $M(p)$).

\section{Property Evaluation}
\label{sec:prop}

The symbolic (time coverage) reachability graph contains several exploitable information.

%DEADLOCK
The tool recognizes deadlocks even if they are topologically hidden by the presence of outgoing edges. In fact if all the outgoing edges have a white tail, it is still possible that  a proper subset of the corresponding symbolic state is composed by deadlock marking. In the running example however no deadlock marking is reachable.

Disregarding time specification (i.e., considering only the number of tokens distributed over places),
the graph nodes exactly identify all the reachable (topological) markings: if a marking matches a symbolic node then there exists at least one path from the initial state to such a marking,
conversely if a marking matches no symbolic nodes, it is not reachable.
It is thereby possible to verify P-invariants from a specified marking.
%%La frase seguente si puÚ togliere
In case of finite graph, it is possible to answer questions about maximum (minimum) number of tokens in some (combinations) of places.

In general, due to \emph{TA} introduction, the set of ordinary markings covered (Definition~\ref{state-cov}) by the states of the symbolic graph built
from a TB net is a superset of the reachable ordinary markings of the TB net.
Given a symbolic state $S = \langle M,C \rangle$ , each numerical substitution of $\{T_i\}$ symbols occurring in $M$ and satisfying $C$ corresponds to the projection of reachable ordinary states.
If we are interested in checking timing relations between token's timestamps on the states of the graph we
can get three different answers upon graph inspection: a positive one (e.g., there exists a node that satisfies the condition), a negative one (e.g., no nodes satisfy the condition), or a possibly positive.
For example, if we are looking for a state where a token in place \emph{Flame}
carries on a timestamp greater than the one in place \emph{IGNITION\_PHASE\_S},
state $S9$ provides us with a positive answer. Instead, if we are checking whether
places \emph{Gas} and \emph{Ignition} can ever hold the same timestamp the answer is may be
(the presence of \emph{TA} in either places \emph{covers} that condition).
%the graph permits to define at least a necessary condition

As for timing relations between token's timestamps in different markings, or between firing times in a transition
firing sequence, the symbolic graph permits identifying critical paths by combining the information on edges.
In particular, conservative bounds can be established. In the case they are not enough to exclude incorrect timing behaviors, it is possible to carry out a more accurate analysis by rebuilding a portion of the graph, retracing some critical paths and reintroducing absolute time references. For example, looking at the time information on edges, it is possible to establish that state $S10$ is not reachable from $S0$ in less than 1.7 time units.
We cannot directly infer that $S10$ is reachable in exactly 1.7 time units.

\begin{figure}[htbf]
\centering
\includegraphics[width=0.4\columnwidth]{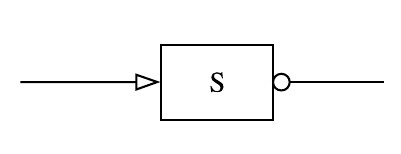}
\caption{Critical case for path feasibility.}
\label{fig:EEpath}
\end{figure}
 
Concerning feasibility of firing sequences (Lemma~\ref{lem:cover}), the symbolic graph expresses all the possibilities 
(an ordinary firing sequence is matched by any firing sequence on the graph). 
A possible critical situation is a white-arrow edge (meaning that we reach only a subset of the target state) is followed by a white-tail edge as shown in Fig.~\ref{fig:EEpath} (meaning that the transition is enabled only in a subset of the ordinary states represented by the node). In this case there is still the possibility that this path actually is not feasible.
Also such critical paths could be retraced.
Let us stress (back to the reachability problem) that by construction, for every node on the graph there exists a path from the initial state to such a node formed exclusively by black-arrow edges.

The available tool's evaluation component is still very simple, its integration with some existing model checking engines is currently under investigation. However it already permits examining the input graph looking for interesting properties on topological definition of markings:
\begin{itemize}
\item existence of a state with a marking satisfying a constraint (i.e., a boolean combination of condition on the number of tokens in places)
\item maximum (minimum) value of an expression involving the number of tokens in places (possibly restricting the evaluation to markings satisfying a given constraint)
\end{itemize}

\section{Tool Architecture}
\label{sec:architecture}

\begin{figure}[htbf]
\centering
\includegraphics[width=\columnwidth]{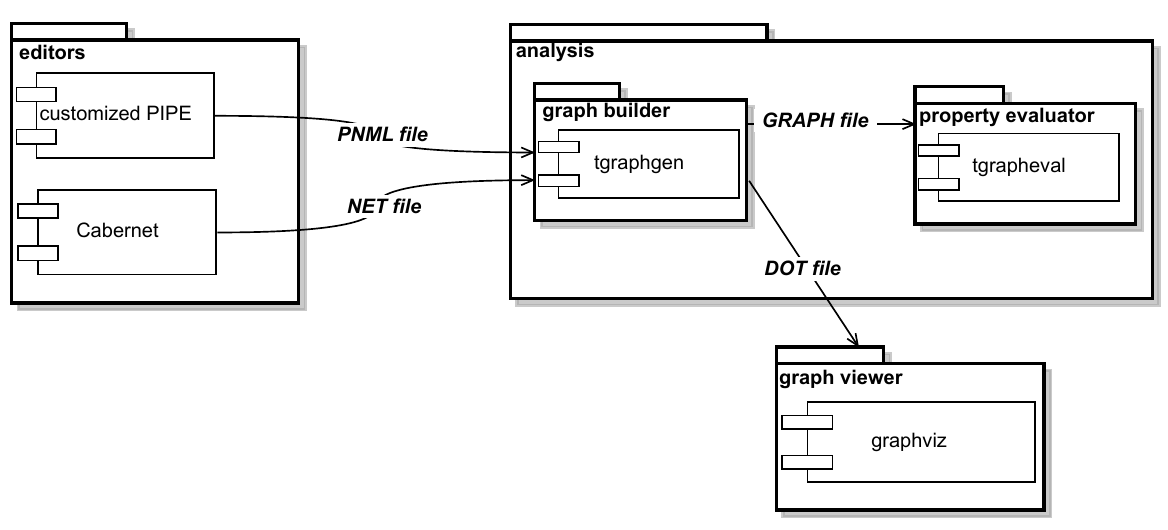}
\caption{Reference architecture.}
\label{fig:architecture}
\end{figure}

The analysis technique described in this paper has been implemented as a command line tool written in Java. The tool architecture depicted in Fig.~\ref{fig:architecture} presents the various components that communicate by means of files.
The \emph{tgraphgen} module receives as input a Time Basic Petri net (either in the legacy file format used by the Cabernet tool, or in a PNML  format generated, for example, by a customized version of \textsc{PIPE2} open source tool\cite{Dingle09}). It generates as outputs the graph in binary format (used by the property verification module $tgrapheval$), and in an annotated DOT text format (used by the \textsc{GraphViz} tool).
The tool is also integrated as an analysis module in the customized \textsc{PIPE2} open source tool. 
That will permit accessing all the functions by means of menu, and exploiting in an integrated environment consolidated structural analysis algorithms for the verification of the untimed part of TB nets (e.g., P/T nets invariant analysis).
Both the command line tool and the customized version of \textsc{PIPE2} are available for download at \url{http://camilli.di.unimi.it/graphgen}, together with a brief user guide and some running examples. 

%\begin{figure}[htbf]
%\centering
%\includegraphics[width=12cm]{PIPEscreenshot}
%\caption{reference architecture}
%\label{fig:PIPE}
%\end{figure}

\section{Use Case and Comparison with other tools}
\label{sec:comparison}
\begin{figure*}[htdb]
\centering
\includegraphics[width=\textwidth]{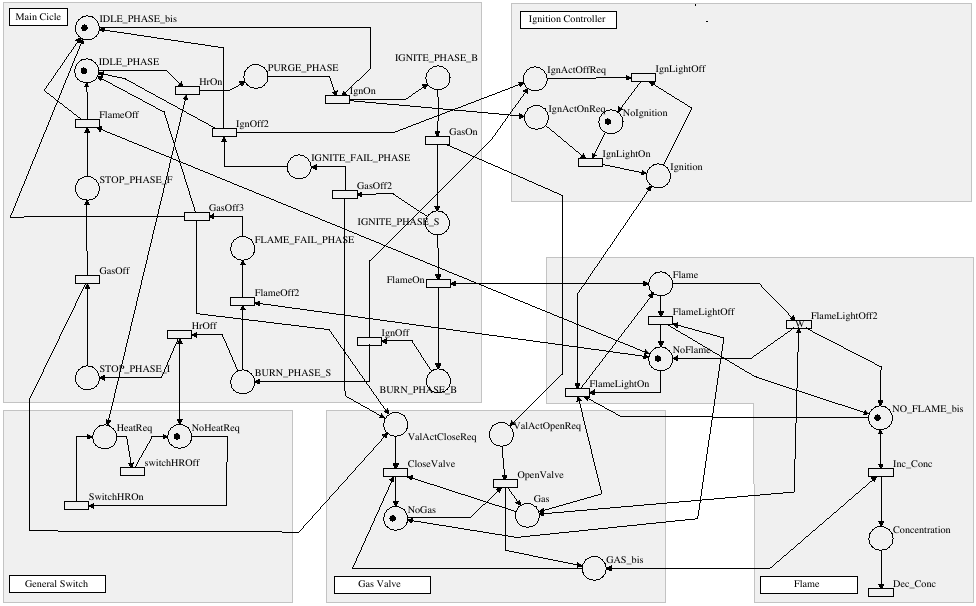}
{\scriptsize
\[
\begin{array}{llll}
\textbf{Initial marking:} &  \multicolumn{3}{l}{IDLE\_PHASE\{T_0\},\,IDLE\_PHASE\_bis\{T_0\},\,NoIgnition\{T_0\}, }\\
  & \multicolumn{3}{l}{NoHeatReq\{T_0\}, \, NoGas\{T_0\}, \, NoFlame\{T_0\},\,NO\_FLAME\_bis\{T_0\} } \\
 \textbf{Initial constraint:} & 0 \le T_0 \le 10 \\
%\textbf{Initial marking:} & \textrm{all tokens in the figure have time }  T_0  & \textbf{Initial constraint:} & 0 \le T_0 \le 10 \\
 & \\
\textbf{Time-Functions:}\\
\textbf{HrOn} & \multicolumn{3}{l}{ [ IDLE\_PHASE + 0.01, max (\{IDLE\_PHASE + 0.01 , HeatReq + 0.1\}) ] }\\
\textbf{HrOff} &  \multicolumn{3}{l}{[ BURN\_PHASE\_S + 0.01, } \\
& \multicolumn{3}{l}{max ( \{BURN\_PHASE\_S + 0.01 , NoHeatReq + 0.1 \}) ]} \\
\textbf{IgnOn} &  \multicolumn{3}{l}{[ max(\{PURGE\_PHASE + 0.01,IDLE\_PHASE\_bis + 30\}),}\\
& \multicolumn{3}{l}{max(\{PURGE\_PHASE + 0.01,IDLE\_PHASE\_bis + 30\}) ]} \\
\textbf{CloseValve} &  [ ValActCloseReq + 0.2, ValActCloseReq + 0.2 ] \\
\textbf{OpenValve} &  [ ValActOpenReq + 0.2, ValActOpenReq + 0.2 ] \\
\textbf{FlameOff} &  \multicolumn{3}{l}{[ STOP\_PHASE\_F + 0.01, }\\
& \multicolumn{3}{l}{max ( \{STOP\_PHASE\_F+0.01, NoFlame+0.1 \}) ] } \\
\textbf{FlameOff2} &  \multicolumn{3}{l}{[ BURN\_PHASE\_S + 0.01, } \\
& \multicolumn{3}{l}{max ( \{BURN\_PHASE\_S+0.01 , NoFlame+0.1 \}) ] } \\
\textbf{FlameOn} &  \multicolumn{3}{l}{[ IGNITE\_PHASE\_S + 0.01, }\\
& \multicolumn{3}{l}{max ( \{BURN\_PHASE\_S+0.01 , NoFlame+0.1 \}) ] } \\
\textbf{IgnLightOn} &  \multicolumn{3}{l}{[ IgnActOnReq + 0.2, IgnActOnReq + 0.2 ]} \\
\textbf{IgnLightOff} &  \multicolumn{3}{l}{[ IgnActOffReq + 0.2, IgnActOffReq + 0.2 ]}  \\
\textbf{FlameLightOn} &  \multicolumn{3}{l}{[ max (\{Gas , Ignition \}) + 0.5, max (\{Gas , Ignition \}) + 0.5 ]}\\
\textbf{FlameLightOff} &  [ enab, NoGas + 0.1 ] \\
\textbf{FlameLightOff2} &  [ enab, enab+100 ] \\
\textbf{GasOn} &  [ enab + 0.01, enab + 0.1 ] \\
\textbf{GasOff} &  [ enab + 0.01, enab + 0.1 ] \\
\textbf{GasOff2} &  [ enab + 2, enab + 2 ] \\
\textbf{GasOff3} &  [ enab + 0.01, enab + 0.1 ] \\

\textbf{IgnOff} &  [ enab + 0.01, enab + 0.1 ] \\
\textbf{IgnOff2} &  [ enab + 0.01, enab + 0.1 ] \\

\textbf{SwitchHROn} &  [ enab, enab + 10 ] \\
\textbf{switchHROff} &  [ enab + 120, enab + 120 ] \\
\textbf{Inc\_Conc} &  [ enab+0.1, enab+0.1 ] \\
\textbf{Dec\_Conc} &  [ enab+30, enab+30 ] \\
\end{array}
\]
}
\caption{Use case net: gas burner.}
\label{fig:gasburner}
\end{figure*}

In order to make a comparison with the available analysis techniques and tools for TB nets, 
we consider now the complete gas burner example analyzed in \cite{IPTES-PDM41}, also reported in Fig.~\ref{fig:gasburner}) for completeness.

The main critical parameter of the system was identified in the concentration value of unburned gas. With the old analyzers  
it was only possible to do an approximate analysis, by verifying the safety requirement within a fixed time threshold \cite{IPTES-PDM41}, 
%or by building a small part of the reachability tree able to invalidate the property \cite{Calzolari}. 
or by empirically guiding the construction of a portion of the reachability tree looking for a state invalidating the property \cite{Calzolari}.
These techniques were only able to verify the unsatisfiability of the time bounded safety property by ending the construction of the tree after reaching a state with a  concentration exceeding a critical value (i.e., according the specification, one second of unburned gas).
A significant improvement is that our technique computes the graph 
representing the complete behavior of the system, and thus for example permits calculating the actual concentration upper bound.

\begin{figure}[htbf]
\centering
\includegraphics[width=.8\columnwidth]{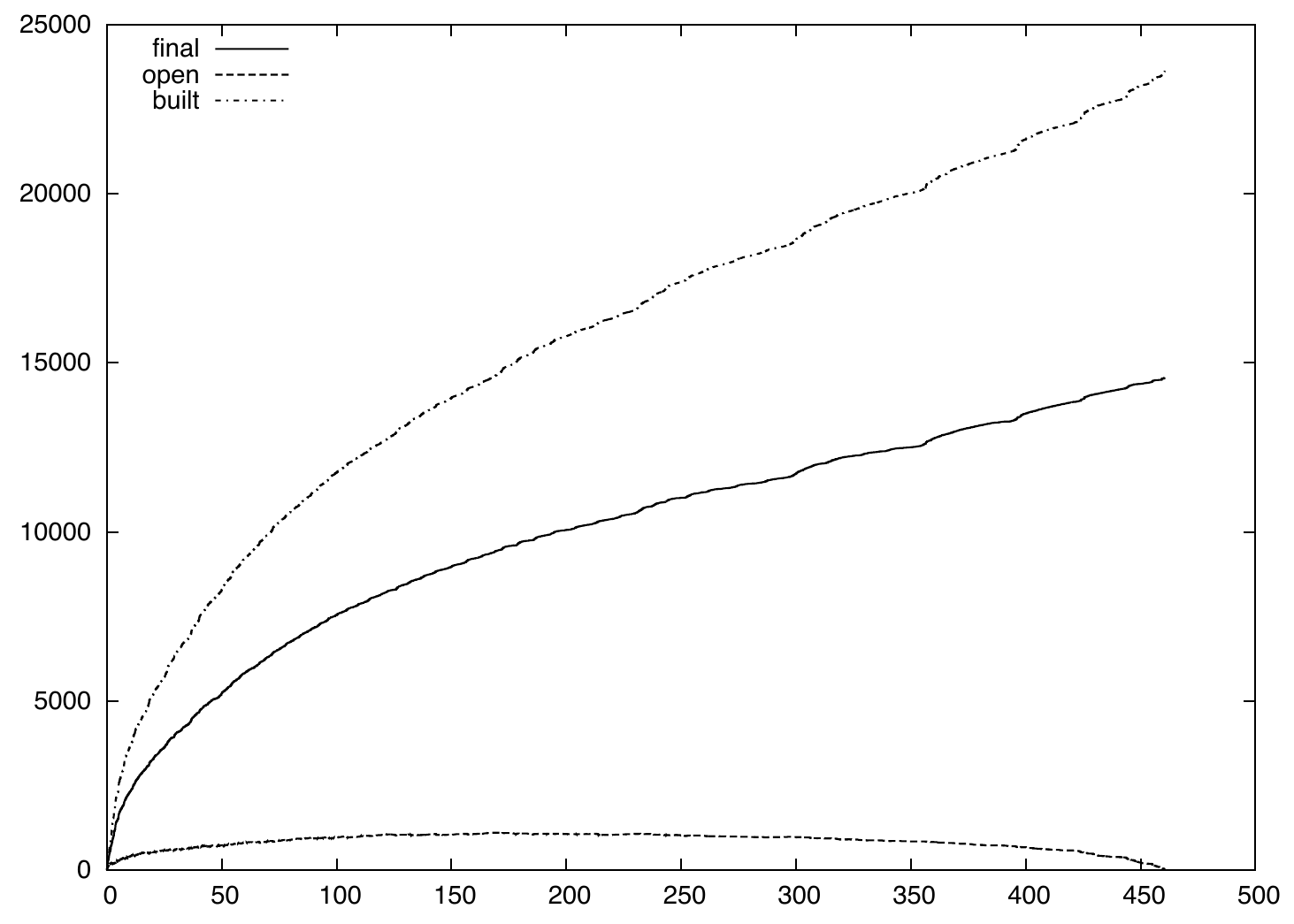}
\caption{State creation advancement.}
\label{fig:01graph}
\end{figure}

Table~\ref{tab:results} reports the outcomes of the analysis on the use case. In particular the considered parameter has been measured with three versions of the net. They differ in the time granularity used for the unburned gas process, i.e., the time function of the transition $Inc\_Conc$. The first thing to note is however that the analysis result is coherent in the various situations, identifying the maximum amount of unburned gas as corresponding to a leaking period of two seconds.

The test has been performed on a Toshiba Notebook with 2.4Ghz Intel Core 2 Duo processor and 4GB of memory. The operating system is Ubuntu 10.10 and the Java Virtual Machine is OpenJDK IcedTea6 1.9.5.

On the table we report also the number of states of the final reduced graph against the overall number of states
generated by the algorithm, and the execution times.

In Fig.~\ref{fig:01graph} some profiling data -- relating the 0.1 time granularity version of the model --
are presented.
On the x axis there is the execution time expressed in minutes, on the y axis there are the number of built nodes, of
reduced (final) nodes, and of nodes ready to be processed, respectively.
This picture is important for two reasons: first it shows that the performance degradation of state construction process is very small (the number of states created is pretty much constant in time after an initial burst); second, it supports the idea that a parallel (distributed) version of the graph builder, introduced in \cite{Camilli12,Camilli12-2,Camilli13}
%currently under development, 
should substantially improve the performances (the front of expansion remaining consistently wide).
%NOTA: questo dipende pi che dalla tecnica ... dal modello... boh... lo diciamo

\begin{table}[htdp]
\caption{Use case analysis results.}
\begin{center}
\begin{tabular}{@{} |c|c|c|c|c| @{}}
\hline \hline
$Inc\_Conc$ gran. & max(Conc) & \#  [final/built] states & exec. time\\
\hline \hline
0.5 &  4 & 865/1217  & $\approx 75 secs$ \\
\hline
0.25 &  8 & 2233/2983 & $\approx 400 secs$ \\
\hline
0.1 &  20 &  14563/23635  & $\approx 7.5 hrs$ \\
\hline
\end{tabular}
\end{center}
\label{tab:results}
\end{table}%

%\begin{figure}[htbf]
%\centering
%\includegraphics[angle=90,width=5.5cm]{gas_burner_conc05}
%\caption{Use Case reachability graph}
%\label{fig:graph05}
%\end{figure}

\section{Conclusion and future works}
\label{sec:conclusion}

The analysis technique presented in this paper overtakes the existing available
analysis technique for Time Basic Nets (a very expressive timed version of Petri nets) because it permits the building of a sort of (symbolic) time-coverage reachability graph keeping interesting timing properties of the nets.
In particular the introduction of the concept of \emph{time anonymous} timestamps, allows for a
major factorization of symbolic states.
An extension of the technique that further exploits the time anonymous concept in order to deal with topologically unbounded nets (by means of a coverage of \emph{TA} tokens, i.e., a sort of $\omega_{\textrm{\tiny{TA}}}$) is under definition.

\bibliographystyle{plain}
\bibliography{PN2011}

\begin{thebibliography}{10}

\bibitem{Ravn93}
A.~P. Atlee and H.~Gannon.
\newblock Specifying and verifying requirements of real-time systems.
\newblock {\em IEEE Trans. Softw. Eng.}, 19:41--55, January 1993.

\bibitem{Camilli12}
Carlo Bellettini, Matteo Camilli, Lorenzo Capra, and Mattia Monga.
\newblock Symbolic state space exploration of {RT} systems in the cloud.
\newblock In {\em Symbolic and Numeric Algorithms for Scientific Computing},
  SYNASC 2012, pages 295--302, Los Alamitos, CA, USA, 2012. IEEE CS Press.

\bibitem{Camilli13}
Carlo Bellettini, Matteo Camilli, Lorenzo Capra, and Mattia Monga.
\newblock Mardigras: Simplified building of reachability graphs on large
  clusters.
\newblock In ParoshAziz Abdulla and Igor Potapov, editors, {\em Reachability
  Problems}, volume 8169 of {\em LNCS}, pages 83--95. Springer Berlin
  Heidelberg, 2013.

\bibitem{Bellettini11}
Carlo Bellettini and Lorenzo Capra.
\newblock Reachability analysis of time basic petri nets: A time coverage
  approach.
\newblock In {\em Proceedings of the 2011 13th International Symposium on
  Symbolic and Numeric Algorithms for Scientific Computing}, SYNASC '11, pages
  110--117, Washington, DC, USA, 2011. IEEE Computer Society.

\bibitem{Merlot93}
Carlo Bellettini, Miguel Felder, and Mauro Pezz\`{e}.
\newblock Merlot: a tool for analysis of real-time specifications.
\newblock In {\em Proceedings of the 7th international workshop on Software
  specification and design}, IWSSD '93, pages 110--119, Los Alamitos, CA, USA,
  1993. IEEE Computer Society Press.

\bibitem{IPTES-PDM41}
Carlo Bellettini, Miguel Felder, and Mauro Pezz{\`e}.
\newblock A tool for analysing high-level timed petri nets.
\newblock IPTES Esprit Project 5570 PDM-41, Politecnico di Milano, September
  1993.

\bibitem{Berthomieu91}
Bernard Berthomieu and Michel Diaz.
\newblock Modeling and verification of time dependent systems using time petri
  nets.
\newblock {\em IEEE Trans. Softw. Eng.}, 17:259--273, March 1991.

\bibitem{Calzolari}
F.~Calzolari and M.~Pezz\`{e}.
\newblock Property decomposition to speed up analysis.
\newblock {\em Real-Time Systems, Euromicro Conference on}, 0:147, 1995.

\bibitem{Camilli12-2}
Matteo Camilli.
\newblock {P}etri nets state space analysis in the cloud.
\newblock In {\em Proceedings of the 2012 International Conference on Software
  Engineering}, ICSE 2012, pages 1638--1640, Piscataway, NJ, USA, 2012. IEEE
  Press.

\bibitem{Dingle09}
Nicholas~J. Dingle, William~J. Knottenbelt, and Tamas Suto.
\newblock Pipe2: A tool for the performance evaluation of generalised
  stochastic petri nets.
\newblock {\em SIGMETRICS Perform. Eval. Rev.}, 36(4):34--39, March 2009.

\bibitem{UnifiedWay91}
Carlo Ghezzi, Dino Mandrioli, Sandro Morasca, and Mauro Pezz\`{e}.
\newblock A unified high-level petri net formalism for time-critical systems.
\newblock {\em IEEE Trans. Softw. Eng.}, 17:160--172, February 1991.

\bibitem{Ghezzi94}
Carlo Ghezzi, Sandro Morasca, and Mauro Pezz\`{e}.
\newblock Validating timing requirements for time basic net specifications.
\newblock {\em J. Syst. Softw.}, 27:97--117, November 1994.

\bibitem{Cabernet93}
Carlo Ghezzi and Mauro Pezz\`{e}.
\newblock Towards extensible graphical formalisms.
\newblock In {\em Proceedings of the 7th international workshop on Software
  specification and design}, IWSSD '93, pages 69--77, Los Alamitos, CA, USA,
  1993. IEEE Computer Society Press.

\bibitem{graphviz}
http://www.graphviz.org/.
\newblock Graphviz - graph visualization software.

\bibitem{Hudak2010}
A.N. Kovacs and S.~Hudak.
\newblock Time semantics in time basic nets.
\newblock In {\em Applied Machine Intelligence and Informatics (SAMI), 2010
  IEEE 8th International Symposium on}, pages 315 --319, January 2010.

\end{thebibliography}

%\onecolumn
%\appendix

%In this appendix, we present an additional figure for the reviewers, showing an annotated version of the sample built graph (already shown in figure~\ref{fig:graph}) with the definition of symbolic states not commented in the article.
%The tool is not currently available online, but upon  reviewers' request, it is possible to send it to them.

%\begin{figure*}[htbf]
%\centering
%\includegraphics[width=0.8\textwidth]{fig2tot}
%\caption{Annotated version of  Fig.\ref{fig:graph}.}
%\label{fig:graphannotated}
%\end{figure*}

\end{document}